\documentclass{cernrep} 
\usepackage{texnames}
\usepackage[T1]{fontenc}
\usepackage{amsmath,amssymb,bm,slashed,array}
\usepackage{amsfonts}
\usepackage{graphicx}
\usepackage{ifpdf}
\usepackage{color}
\usepackage{multirow}
\usepackage{booktabs}
\usepackage{subfigure}
\usepackage{wasysym}

\definecolor{nicered}{rgb}{0.7,0.1,0.1}
\definecolor{nicegreen}{rgb}{0.1,0.5,0.1}
\definecolor{red}{rgb}{1.0, 0, 0}

\allowdisplaybreaks

\newcommand{\bra}[1]{\ensuremath{\langle #1 |}}   
\newcommand{\ket}[1]{\ensuremath{| #1 \rangle}}   

\newcommand{\beq}{\begin{equation}}
\newcommand{\eeq}{\end{equation}}
\newcommand{\bea}{\begin{eqnarray}}
\newcommand{\eea}{\end{eqnarray}}

\begin{document}

\title{Flavour Physics and CP Violation}

\author{J.~F.~Kamenik}

\institute{J. Stefan Institute, Ljubljana, Slovenia\\
Department of Physics, University of Ljubljana, Ljubljana, Slovenia}

\maketitle

\begin{abstract}
These notes represent a summary of  three lectures on flavour and CP violation, given at the CERN's European School of High Energy Physics in 2014. They cover flavour physics within the standard model, phenomenology of CP violation in meson mixing and decays, as well as constraints of flavour observables on physics beyond the standard model. In preparing the lectures (and consequently this summary) I drew heavily from several existing excellent and exhaustive sets of lecture notes and reviews on flavour physics and CP violation~\cite{refs}. The reader is encouraged to consult those as well as the original literature for a more detailed study.
\end{abstract}

\section{What is flavour?}

In the standard model (SM) the basic constituents of matter are excitations of fermionic fields with spin $1/2$. In this context {\it matter flavours} refers to several copies of the same gauge representation. Under the unbroken SM gauge group $SU(3)_c \times U(1)_{\rm EM}$ these are
\begin{itemize}
\item up-type quarks: $(3)_{2/3}$ : $u,~c,~t$,
\item down-type quarks: $(3)_{-1/3}$ : $d,~s,~b$,
\item chrged leptons: $(1)_{-1}$ : $e,~\mu,~\tau$,
\item neutrinos: $(1)_{0}$ : $\nu_1,~\nu_2,~\nu_3$,
\end{itemize}
where the colour representations are given in the brackets, while the electric charges are written as subscripts. The different flavours of the same gauge representation differ only in their masses.

Ordinary matter is essentially made up of  the first generation: $u$ and $d$ quarks are bound within protons and neutrons, while the electrons form atoms; finally ``electron neutrinos", which are an admixture of $\nu_{1,2,3}$, are produced in reactions inside stars. Second and third generation families are produced only in high-energy particle collisions. They all decay via weak interactions into first generation particles. One of the big open questions in fundamental physics is why there are thee almost identical replicas of quarks and leptons and which is the origin of their different masses?

{\it Flavour physics} refers to interactions that distinguish between flavours. Within the SM these are weak and Yukawa (Higgs boson) interactions.

{\it Flavour parameters} are those that carry flavour indices. Within the SM these are the nine masses of charged fermions and four mixing parameters (three angles and one complex CP violating phase).\footnote{Adding Majorana mass terms for neutrinos introduces three additional neutrino masses plus six mixing parameters (three mixing angles and three phases).}

{\it Flavour universal} interactions are those with couplings proportional to the identity in flavour space. Within the SM these are strong and electromagnetic interactions (and also weak interactions in the so-called interactions basis, see below). Such interactions are sometimes also called {\it flavour blind}.

{\it Flavour diagonal} interactions are those whose couplings are diagonal (in the matter mass basis), but not necessarily universal. Within the SM these are the Yukawa interactions of the Higgs boson.

{\it Flavour changing} processes are those where the initial and final flavour-numbers are different (a flavour number is the number of particles with a certain flavour minus the number of anti-particles of the same flavour). We can further specify {\it flavour changing charged currents} which involve both up- and down-type quark flavours or both charged lepton and neutrino flavours. Examples of such processes are the muon decay $\mu^- \to e^- \nu_i \bar \nu_j$ or the muonic charged kaon decay $K^- \to \mu^- \bar \nu_i$ (which corresponds to the quark-level transition $s\bar u \to \mu^- \bar \nu_i$). Within the SM such processes are mediated already by a single $W$ exchange at the tree level (the amplitudes being proportional to the Fermi constant $G_F$). On the other hand, {\it flavour changing neutral currents} (FCNCs) involve either up- or down- type flavours but not both; and/or either charged lepton flavours or neutrino flavours but not both. Examples of such processes are the radiative muon decay $\mu^- \to e^- \gamma$ and the muonic decays of the neutral kaons, $K_L \to \mu^+ \mu^-$ ($s\bar d \to \mu^+ \mu^-$ at the quark level). Within the SM these processes occur at higher orders in the weak expansion (i.e. via loops) and are often {\it highly suppressed}. In connection with flavour changing interactions, one often speaks also of {\it flavour violation}.

\subsection{Why is flavour interesting?}

Flavour physics can discover new physics (NP) or probe it before it is directly observed in high-energy experiments. Historical examples of this include:
\begin{itemize}
\item The smallness of the ratio $\Gamma(K_L \to \mu^+ \mu^-) / \Gamma(K^- \to \mu^- \bar \nu_i)$ lead to the prediction of the charmed quark.
\item Furthermore, the measurement of the mass difference between the two neutral kaons $\Delta m_K \equiv m_{K_L} - m_{K_S}$ lead to the prediction of the charm quark mass. 
\item Similarly, the mass difference between the two neutral $B$ mesons $\Delta m_B \equiv m_{B^0_H} - m_{B^0_L}$ inferred a prediction of the top quark mass almost two decades before top quarks (or more precisely, their decay products) were directly observed in experiments. 
\item Finally, the observation of the CP violating decay $K_L \to \pi^+ \pi^-$ (i.e the measurement of $\epsilon_K$) lead to the prediction of the third generation of matter. 
\end{itemize}

{\it CP violation}: Within the SM there is a single CP violating parameter determining the amount of CP violation in all flavour changing processes. Successful baryogenesis would require new CP violating sources.

Solutions of the electroweak (EW) hierarchy problem (in the form of a quadratic sensitivity of the EW scale to UV physics) require NP to appear at or below the TeV scale. On the other hand, such NP with a generic flavour structure would predict FCNCs orders of magnitude above the observed rates. Conversely, flavour physics can probe NP scales up to $\mathcal O(10^5~{\rm TeV})$. The resulting {\it NP flavour puzzle} refers to the fact that NP at the TeV scale needs to exhibit approximate flavour symmetries.

The SM flavour parameters are both hierarchical (i.e. $m_u \ll m_c \ll m_t$) and mostly very small ($m_{f\neq t} \ll m_{W,Z,h}$)\,. The question whether this points to some unknown underlying flavour dynamics is sometimes called the {\it SM flavour puzzle}.

\section{Flavour in the standard model}

Any (local) quantum field theory model is specified by both (i) symmetries and the pattern of their spontaneous breaking; as well as (ii) representations of fermions and scalars. The SM Lagrangian ($\mathcal L_{\rm SM}$) is thus completely determined by specifying the local (gauge) symmetry $\mathcal G_{\rm local}^{\rm SM} = SU(3)_c \times SU(2)_L \times U(1)_Y$ which is spontaneously broken to $\mathcal G_{\rm local}^{\rm SM} \to SU(3)_c \times U(1)_{\rm EM}$\,; plus the relevant fermionic
\begin{align}
Q_L^i \sim (3,2)_{1/6}\,, & & U_R^i\sim(3,1)_{2/3}\,, & & D_R^i\sim (3,1)_{-1/3}\,, & & L_L^i\sim (1,2)_{-1/2}\,,
\end{align}
(where $i=1,2,3$) and scalar 
 \begin{align}
 \phi & \sim(1,2)_{1/2}\,, & \langle \phi^0 \rangle &\equiv \frac{v}{\sqrt 2} \simeq 174 \rm GeV\,,
 \end{align}
 representations. Above, the first (second) entries in the brackets denote the $SU(3)_c$ ($SU(2)_L$) representations, while the $U(1)_Y$ charges are given in the subscripts. Also, $\langle \ldots\rangle$ denotes a vacuum condensate value. $\mathcal L_{\rm SM}$ can be conveniently split into three parts
 \begin{align}
 \mathcal L_{\rm SM} =  \mathcal L_{\rm kinetic}^{\rm SM} +  \mathcal L_{\rm EWSB}^{\rm SM} +  \mathcal L_{\rm Yukawa}^{\rm SM}\,.
 \end{align} 
 The sum of the gauge-kinetic terms $\mathcal L_{\rm kinetic}^{\rm SM}$ is simple and symmetric. It is completely specified by the SM local symmetry and its matter representations. The three physical parameters associated with this part of the theory are conventionally chosen to be the three gauge couplings ($g_s$, $g$ and $g'$)\,. The EW symmetry breaking (EWSB) part $\mathcal L_{\rm EWSB}^{\rm SM}$ contains two additional parameters. They can be chosen to correspond to $v$  and the physical Higgs boson mass $m_h$. Finally, all flavour dynamics is contained in $\mathcal L_{\rm Yukawa}^{\rm SM}$ which also involves all the SM flavour parameters. 
 
 \subsection{Interaction basis}
 
It is convenient to start our discussion in a flavour basis where all the gauge-kinetic terms are diagonal. This can always be achieved by applying suitable unitary rotations on the matter fields.  In this basis
\begin{align}
\mathcal L_{\rm kinetic}^{\rm SM} &= (D_\mu \phi)^\dagger (D^\mu \phi) + \sum_{i,j = 1,2,3} \sum_{\psi = Q_L,\ldots,E_R} \bar \psi^i i \slashed D \delta^{ij} \psi^j \nonumber \\
&- \frac{1}{4} \sum_{a=1,\ldots,8} G_{\mu\nu}^a G^{a,\mu\nu} - \frac{1}{4} \sum_{a=1,2,3} W_{\mu\nu}^a W^{a,\mu\nu}  - \frac{1}{4} B_{\mu\nu} B^{\mu\nu} \,,
\end{align}
where $G,~W,$ and $B$ denote the field strengths of the $SU(3)_c$, $SU(2)_L$ and $U(1)_Y$ gauge interactions, respectively. The covariant derivatives $D_\mu$ are defined as $D_\mu = \partial_\mu + i g_s G_\mu^a L^a + i g W_\mu^b T^b + i g' B_\mu Y$\,, where $L^a$, $T^a$ and $Y$ denote the $SU(3)_c$, $SU(2)_L$ generators and the $U(1)_Y$ charges, respectively. Note that in this basis, $\mathcal L_{\rm kinetic}^{\rm SM}$ is manifestly flavour universal and CP conserving. Similarly
\begin{align}
\mathcal L^{\rm SM}_{\rm EWSB} = \mu^2 \phi^\dagger \phi - \lambda (\phi^\dagger \phi)^2\,,
\end{align}
is also CP and flavour conserving.\footnote{It is also symmetric under $SO(4)$ rotations of the four real scalar fields $\phi_{1,2,3,4}$ contained in $\phi = (\phi_1 + i \phi_2, \phi_3 + i \phi_4)^T$. This approximate symmetry of the SM is sometimes called the {\it custodial symmetry}.} Thus both $\mathcal L_{\rm kinetic}^{\rm SM} $ and trivially $\mathcal L_{\rm EWSB}^{\rm SM}$ have a large flavour symmetry corresponding to the independent unitary rotations in the flavour space of the five fermionic fields
\begin{align}
\mathcal G_{\rm flavour}^{\rm SM} &= U(3)^5 = SU(3)_q^3 \times SU(3)_\ell^2  \times U(1)^5 \,, \nonumber\\
SU(3)_q^3 &= SU(3)_Q \times SU(3)_U \times SU(3)_D\,,\nonumber\\
SU(3)_\ell^2 &= SU(3)_L \times SU(3)_E \,, \nonumber\\
U(1)^5 &= U(1)_B \times U(1)_L \times U(1)_Y \times U(1)_{\rm PQ} \times U(1)_E\,.
\end{align}
Among the $U(1)$ factors, $U(1)_{B,L}$ are the baryon and lepton number, respectively. $U(1)_Y$ is gauged and broken spontanouesly by $\langle \phi^0 \rangle$\,. On the other hand $U(1)_{\rm PQ}$ can be defined such that only the Higgs and $D^i_R,E^i_R$ are charged under it and with opposite charges. It is thus broken only by the up-quark Yukawas. Finally $U(1)_E$ refers to flavour universal phase rotations of $E^i_R$ alone and is thus broken by the charged lepton Yukawas.

The Yukawa Lagrangian of the SM
\begin{align}
- \mathcal L_{\rm Yukawa}^{\rm SM} = Y_d^{ij} \bar Q_L^i \phi D_R^j + Y_u^{ij} \bar Q_L^i \tilde \phi U_R^j + Y_e^{ij} \bar L^i \phi E_R^j + \rm h.c.\,,
\end{align}
where $\tilde \phi = i \sigma_2 \phi$, is in general flavour dependent (if $Y_f \slashed{\propto} \,\mathrm I$) and CP violating. The pattern of explicit $\mathcal G_{\rm flavour}^{\rm SM}$ breaking by $Y_f \neq 0$ is as follows:
\begin{itemize}
\item $U(1)_E$ is broken by $Y_e \neq 0$\,,
\item $U(1)_{\rm PQ}$ is broken by $Y_u \cdot Y_d \neq 0$ and $Y_u\cdot Y_e \neq 0$\,,
\item $SU(3)_Q \times SU(3)_U \to U(1)_u \times U(1)_c \times U(1)_t$ is due to $Y_u \slashed \propto \, \rm I$\,,
\item $SU(3)_Q \times SU(3)_D \to U(1)_d \times U(1)_s \times U(1)_b$ is due to $Y_d \slashed \propto \, \rm I$\,,
\item the remaining $U(1)$ factors in the quark sector are broken by the fact that $[Y_u , Y_d] \neq 0$ down to $U(1)_B$ \,,
\item finally, $SU(3)_L \times SU(3)_E \to U(1)_e \times U(1)_\mu \times U(1)_\tau$ due to $Y_e \slashed \propto \, \rm I$\,. The remaining factor group also contains the global $U(1)_L$\,.
\end{itemize}
Thus, the global symmetry of the SM in presence of the Yukawas is $\mathcal G_{\rm global}^{\rm SM}(Y_f\neq 0) = U(1)_B \times U(1)_e \times U(1)_\mu \times U(1)_\tau$\,. In this language, {\it flavour physics} refers to interactions which break the $SU(3)_q^3 \times SU(3)_\ell^2$ and are thus {\it flavour violating}.

Commonly, a spurion analysis is useful for parameter counting, identification of suppression factors, and for the idea of minimal flavour violation (MFV)~\cite{mfv}. In this approach we promote the SM Yukawas to non-dynamical fields with well-defined transformation properties under $\mathcal G_{\rm flavour}^{\rm SM}$
\begin{align}
 Y_u & \sim (3,\bar 3,1)_{SU(3)_q^3}\,, & Y_d &\sim (3,1,\bar 3)_{SU(3)_q^3}\,, & Y_e & \sim (3,\bar 3)_{SU(3)_\ell^2}\,.
\end{align}
In the following we will focus on the quark sector. 

\subsection{Counting the standard model quark flavour parameters}

The flavour symmetry breaking pattern described above is useful in counting the number of physical flavour parameters in the theory. In particular:
\begin{enumerate}
\item Consider a theory with a global symmetry group $\mathcal G_f$ with $N_{\rm total}$ generators.
\item Add interactions with $N_{\rm general}$ parameters, breaking $\mathcal G_f \to \mathcal H_f$ with $N_{\rm total} - N_{\rm broken}$ generators.
\item Then the $N_{\rm broken}$ generators can be used to rotate away $N_{\rm broken}$ number of symmetry breaking parameters.
\item The number of remaining physical parameters is thus $N_{\rm physical} = N_{\rm general} - N_{\rm broken}$\,.
\end{enumerate}
We can apply this recipe to the SM breaking of $U(3)_Q \times U(3)_U \times U(3)_D \to U(1)_B$. In this case the three $U(3)$ group rotations are described by unitary $3\times 3$ matrices containing three real angles and six phases each. Thus schematically $N_{\rm total} = 3\times (3 + 6 i )$\,. Consequently $N_{\rm broken} = N_{\rm total} - 1 i = 9 + 17 i$\,. The two quark Yukawas are general $3\times 3$ matrices containing nine complex parameters ($N_{\rm general} = 2 \times (9 + 9 i)$). Finally, the number of physical parameters is $N_{\rm physical} = N_{\rm general} - N_{\rm broken} = 9 + 1 i$, representing six quark masses, three mixing angles and a single CP violating phase.

\subsection{Discrete symmetries of the standard model}

Any local Lorentz invariant quantum field theory conserves CPT~\cite{cpt}. It follows that in these theories (including the SM) T violation equals CP violation. There is no reason, a priori, for C, P and CP to be related to flavour physics. However, in the SM (and apparently in Nature) this is so. In the SM C and P are violated maximally: left-handed and right-handed fermion fields furnish different gauge representations, while C and P both change the chirality of fermion fields. This maximal C and P violation within the SM is also independent of the values of the SM parameters. On the other hand, the CP violation within the SM does depend on the (Yukawa) parameters. The hermiticity of the Lagrangian namely implies
\begin{align}
Y_{ij} \bar \psi^i_L \phi \psi_R^j + Y_{ij}^* \bar \psi_R^j \phi^\dagger \psi_L^i \overset{\rm CP}{\to} Y_{ij} \bar \psi_R^j \phi^\dagger \psi_L^i + Y_{ij}^* \bar \psi_L^i \phi \psi_R^j\,.
\end{align}
Thus, the Yukawa Lagrangian will be CP symmetric if $Y_{ij} = Y_{ij}^*$. More precisely, the requirement for CP conservation can be written in terms of the Jarlskog invariant ($J$)~\cite{Jarlskog:1985ht} as
\begin{align}
J\equiv {\rm Im}[{\rm det} ( Y_d Y_d^\dagger,Y_u Y_u^\dagger  )]  = 0\,.
\end{align}

\subsection{Mass basis}

Upon replacing ${\rm Re}(\phi^0) \to (v+h)/\sqrt{2}$, Yukawa interactions give rise to fermion mass matrices 
\begin{align}
M_q = \frac{v}{\sqrt 2} Y_q\,.
\end{align}
The {\it mass bassis} corresponds, by definition, to diagonal mass matrices. The unitary transformations between any two bases which leave the gauge-kinetic terms invariant are
\begin{align}
Q_L &\to V_Q Q_L\,, & U_R & \to V_U U_R\,, & D_R & \to V_D D_R \,. 
\end{align}
The Yukawa matrices on the other hand transform as
\begin{align}
Y_u & \to V_Q Y_u V_U^\dagger \,, & Y_d & \to V_Q Y_d V_D^\dagger \,.
\end{align}
The diagonalization of $M_Q$ requires bi-unitary transformations
\begin{align}
V_Q^u M_u V_U^\dagger & = M_u^{\rm diag} = \frac{v}{\sqrt 2} \lambda_u\,; & \lambda_u & = {\rm diag} (y_u, y_c, y_t)\,,\nonumber \\
V_Q^d M_d V_D^\dagger & = M_d^{\rm diag} = \frac{v}{\sqrt 2} \lambda_d\,; & \lambda_d & = {\rm diag} (y_d, y_s, y_b)\,.
\end{align}
While $V_{U,D}$ are unphysical (they leave the gauge-kinetic terms invariant), $V_Q^{u,d}$ produce a physical effect. In particular, since $[M_u,M_d] \neq 0$, a nontrivial mixing matrix $V_Q^u V_Q^{d\dagger} \equiv V_{\rm CKM} \neq \rm 1$  (due to Cabibbo, Kobayashi and Maskawa~\cite{ckm}) modifies the charged weak gauge interactions. The resulting SM flavour Lagrangian in the mass basis is thus
\begin{align}
\mathcal L_m^F & = \left(  \bar q_i \slashed D q^j \delta_{ij} \right)_{\rm NC} + \frac{g}{\sqrt 2} \bar u_L^i \slashed W^+ V^{ij}_{\rm CKM} d_L^j + \bar u_L^i \lambda^{ij}_u u_R^j \left( \frac{v+h}{\sqrt 2}\right) + \bar d_L^i \lambda^{ij}_d d_R^j \left( \frac{v+h}{\sqrt 2}\right)  + \rm h.c.\,,
\end{align}
where $(u^i_L,d^i_L) \equiv Q_L^T$ and NC refers to neutral currents (interactions with gluons, the photon and the Z boson). 

\section{Testing the CKM description of flavour}

Let us recap the main features of quark flavour conversion in the SM: (i) it only proceeds via the three CKM angles; (ii) is mediated by charged current electroweak interactions; and (iii) these charged current interactions involve exclusively left-handed fermion fields.

\subsection{Parametrisation of the CKM matrix}

We start by fixing the permutation of quark generations via mass ordering. The resulting CKM matrix has the form
\begin{equation}
V_{\rm CKM}  =  
\begin{pmatrix}  
V_{ud} & V_{us} & V_{ub} \\
V_{cd} & V_{cs} & V_{cb} \\
V_{td} & V_{ts} & V_{tb} 
\end{pmatrix}
\,.
\end{equation} 
Experimentally, $V_{\rm CKM}$ exhbits a strong hierarchical pattern in off-diagonal elements~\cite{pdg}
\begin{align}
|V_{ud}| &\simeq |V_{cs}| \simeq |V_{tb}| \simeq 1\,, & |V_{us}| &\simeq |V_{cd}| \simeq 0.22\,, \nonumber\\
 |V_{cb}| & \simeq |V_{ts}| \simeq 4\times 10^{-2}\,, & |V_{ub}| & \simeq |V_{td}| \simeq 5 \times 10^{-3}\,.
\end{align}
Such structure can be made explicit in the {\it Wolfenstein expansion}~\cite{Wolfenstein:1983yz} in $\lambda \equiv |V_{us}| \simeq 0.22$
\begin{equation}
V_{\rm CKM}  =  
\begin{pmatrix}  
1-\frac{\lambda^2}{2} & \lambda & A \lambda^3 (\rho - i\eta) \\
-\lambda & 1-\frac{\lambda^2}{2} & A\lambda^2 \\
A \lambda^3(1-\rho-i\eta) & -A\lambda^2 & 1 
\end{pmatrix} + \mathcal O(\lambda^4)\,.
\end{equation} 
The four parameters in this parametrisation $\lambda$, $A$, $\rho$ and $\eta$ can be mapped exactly to the four physical CKM parameters at any order in the $\lambda$ expansion. All are of the order $\mathcal O(0.1-1)$ and the CP violating phase is encoded in the imaginary contribution proportional to $\eta$. Current experimental precision already requires that in phenomenological applications, expansion at least to order $\mathcal O(\lambda^4)$ should be taken into account.

\subsection{Unitarity of the CKM}

Being a unitary matrix, one can derive unitarity conditions on the rows and columns of the CKM matrix, in particular
\begin{align}
\sum_k V_{ik}^* V_{jk} &= \delta_{ij}\,,  & \sum_k V_{ki}^* V_{kj} &= \delta_{ij}\,.
\end{align}
Phenomenologically, the most interesting condition applies for $i=1$ and $j=3$
\begin{equation}
V_{ud} V_{ub}^* + V_{cd} V_{cb}^* + V_{td} V_{tb}^* = 0\,,
\end{equation}
simply because all the three terms on the left hand side are of the same order in $\lambda$. The equation defines a triangle in the complex plane. Normalizing one of the sides to unity
\begin{equation}
\frac{V_{ud} V_{ub}^*}{V_{cd} V_{cb}^* } + \frac{V_{td} V_{tb}^*}{V_{cd} V_{cb}^* } + 1 = 0\,,
\end{equation}
one can re-express it in terms of the Wolfenstein parameters (up to $\mathcal O(\lambda^5)$)
\begin{equation}
[\bar \rho + i \bar \eta] + [(1-\bar \rho) - i\bar \eta] + 1 = 0\,,
\end{equation}
where $\bar \rho = \rho (1 - \lambda^2/2) + \mathcal O(\lambda^4)$ and $\bar \eta = \eta (1 - \lambda^2/2) + \mathcal O(\lambda^4)$\,. The angles (denoted by $\alpha,~ \beta$ and $\gamma$ in Fig.~\ref{fig:CKM}) and sides of this triangle are invariant under phase transformations of quark fields and are observable quantities. 

\subsection{Self consistency of the CKM assumption}

\begin{figure}
\centering
\includegraphics[height=300pt]{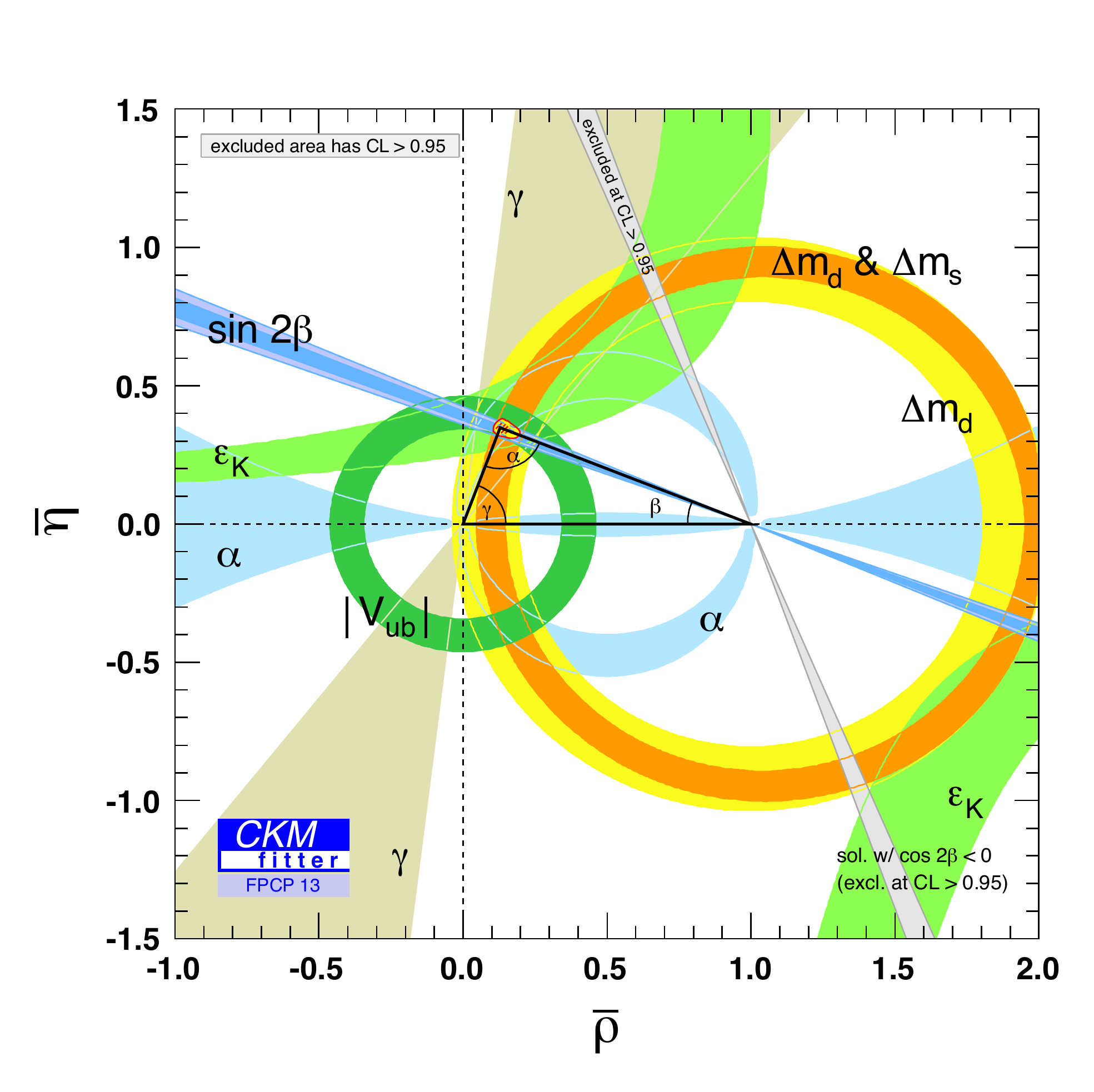}
\caption{Result of the SM CKM fit projected onto the $\bar \rho - \bar \eta$ plane, as obtained by the CKMfitter~\cite{ckmfitter} collaboration. Shown shaded are the 95\% C.L. regions selected by the given observables.}
\label{fig:CKM}
\end{figure}

The CKM description of quark flavour conversion has been tested experimentally to great precision. In particular
\begin{itemize}
\item $|V_{us}|~(\lambda)$ can be extracted from the semileptonic kaon decay $K\to \pi \ell \nu$  with a precision of three per-mille: $\lambda = 0.2253(9)$~\cite{pdg}\,.
\item $|V_{cb}|~(A)$ can be determined from  semileptonic $B$ meson decay width measurements $B \to X_c \ell \nu$ to a precision of two percent: $A=0.822(12)$~\cite{pdg,ckmfitter}\,.
\item Then, $|V_{ub}| \propto \sqrt{\bar\rho^2 + \bar\eta^2}$ can be extracted using  charmless semileptonic decays of $B$ mesons $B\to X_u \ell\nu$\,.
\item The time-dependent CP asymmetry in the decay $B \to \psi K_S$ ($S_{\psi K_S} \simeq \sin 2\beta = 2\bar \eta (1-\bar \rho) / [(1-\bar\rho)^2 + \bar\eta^2]$) has been measured to great precision at the $B$ factory experiments Belle and BaBar.
\item The rates $B\to D K$ decays depend on the phase $\exp(i\gamma) = (\rho + i\eta)/(\rho^2 + \eta^2)$\,.
\item Similarly, the rates of $B\to \pi\pi,\rho\pi,\rho\rho$ depend on the angle $\alpha = \pi - \beta - \gamma$\,.
\item The ratio of neutral $B$ and $B_s$ meson mass diferences $\Delta m_d / \Delta m_s \propto |V_{td}/V_{ts}|^2 = \lambda^2 \left[ (1-\bar\rho)^2 + \bar\eta^2 \right]$ exhibits another non-trivial constraint in the $(\bar \rho,\bar \eta)$ plane. 
\item Finally, CP violation in $K\to \pi\pi$ decays ($\epsilon_K$) depends in a complicate way on  $(\bar \rho,\bar \eta)$.
\end{itemize}
Combined, these measurements lead to an impressive agreement with the best fit ranges for $\rho$ and $\eta$ (see also Fig.~\ref{fig:CKM} and Ref.~\cite{utfit})~\cite{ckmfitter}
\begin{align}
\rho &= 0.130 \pm 0.024\,,  & \eta = 0.362 \pm 0.014\,.
\end{align}

Note that $|\eta| \gtrsim |\rho|$ implies that the CKM phase defined in this way is $\mathcal O(1)$\,.
We can also conclude that, very likely, CP violation in flavour changing processes is dominated by the CKM phase and that the Kobayashi-Maskawa mechanism of CP violation is at work. Again one can define a reparametrisation invariant measure of CP violation
\begin{equation}
{\rm Im} [ V_{ij} V_{kj}^* V_{kl} V_{il}^* ] = J_{KM} \sum \epsilon_{ikm} \epsilon_{jln}\,,
\end{equation}
where $J_{KM}  = \lambda^6 A^2 \eta = \mathcal O(10^{-5})$\,. Written in this form it is clear the CP violation in the SM is suppressed by small mixing among the quark generations. The Jarlskog determinant in the SM can then be written compactly as
\begin{equation}
J = J_{KM} \prod_{i>j} \frac{m_i^2-m_j^2}{v^2} = \mathcal O(10^{-22})\,.
\end{equation}
We see that compared to $J_{KM}$, $J$ is further suppressed by the large quark mass hierarchies.

\section{Closer look at CP violation in neutral meson mixing and decays}
For simplicity, we will focus on the neutral $B$ meson sistem with the flavour eigenstates $B^0 \sim \bar b d$\, and $\bar B^0 \sim b \bar d$. Since in general, these are not CP eigenstates, we have
\begin{align}
CP \ket{B^0} &= e^{i\xi_B} \ket{\bar B^0}\,,\nonumber\\
CP \ket{\bar B^0} &= e^{-i\xi_B} \ket{ B^0}\,.
\end{align}
Stating from an initial superposition state at $t=0$ $\ket{\psi(0)} = 
a(0) \ket{B^0} + b(0) \ket{\bar B^0}$, the time evolution of such a system can in general be described as
\begin{align}
\ket{\psi(t)} = a(t) \ket{B^0} + b(t) \ket{\bar B^0} + c_1(t) \ket{f_1} + c_2(t) \ket{f_2} + \ldots\,,
\end{align}
where $f_{1,2,\ldots}$ denote the $B^0$ and $\bar B^0$ decay products. If we are only interested in $a(t)$ and $b(t)$, we can construct an effective description of the time evolution in terms of a non-hermition Hamiltonian
\begin{equation}
\mathcal H = M + i \frac{\Gamma}{2}\,,
\end{equation}
where $M$ and $\Gamma$ are time-independent, Hermitian $2\times 2$ matrices, describing possible oscillations and decays, respectively. The dispersive part $M$ recieves contributions from  off-shell intermediate states, while $\Gamma$ is the absorptive part and given by a sum over possible on-shell intermediate states. The time-evolution is then described by
\begin{equation}
i \frac{d}{dt} \begin{pmatrix}a(t) \\ b(t)\end{pmatrix} = H \begin{pmatrix}a(t) \\ b(t)\end{pmatrix}\,,
\end{equation}
with the eigenvectors $\ket{B_{L,H}} = p_{L,H} \ket{B^0} \pm q_{L,H} \ket{\bar B^0}$\,, and where $|p_{L,H}|^2 + |q_{L,H}|^2 = 1$\,. Imposing CPT, one obtains $M_{11}=M_{22}$, $\Gamma_{11} = \Gamma_{22}$, and consequently $p_L=p_H \equiv p$ and $q_L=q_H\equiv q$\,. If CP is conserved one furthermore obtains that ${\rm Arg}(M_{12}) = {\rm Arg}(\Gamma_{12})$ and thus $|q/p|=1$\,.

Conventionally, on defines the following CP conserving oscillation parameters 
\begin{align}
m & \equiv \frac{M_L + M_H}{2}\,, & \Gamma & \equiv \frac{\Gamma_L + \Gamma_H}{2}\,, \nonumber \\
\Delta m & \equiv M_H - M_L\,, & \Delta \Gamma & \equiv \Gamma_H - \Gamma_L\,,
\end{align}
or equivalently $x\equiv \Delta m/\Gamma$ and $y\equiv \Delta \Gamma / 2 \Gamma$\,.

The time evolution of the neutral meson system can finally be parametrized in terms of states $\ket{B^0(t)}$ corresponding to $\ket{B^0}$ at initial time $t=0$, and $\ket{\bar B^0(t)}$ corresponding to $\ket{\bar B^0}$ at $t=0$
\begin{align}
\ket{B^0(t)} & = g_+(t) \ket{B^0} - \frac{q}{p} g_-(t) \ket{\bar B^0}\,, \nonumber\\
\ket{\bar B^0(t)} & = g_+(t) \ket{\bar B^0} - \frac{q}{p} g_-(t) \ket{ B^0}\,,
\end{align}
where 
\begin{equation}
g_{\pm} \equiv \frac{1}{2} \left( e^{-m_H t - \Gamma_H t/2} \pm e^{-m_L t - \Gamma_L t/2}  \right)\,.
\end{equation}
The decay of the two mass eigenstates to some some final state $f$ after time t is described by the decay amplitudes
\begin{align}
\bra{f} \mathcal H \ket{B^0} \equiv A_f\,, \nonumber\\
\bra{\bar f} \mathcal H \ket{B^0} \equiv A_{\bar f}\,.
\end{align}
The time-dependent decay rates are then given by
\begin{align}
\frac{d\Gamma(\ket{B^0(0)}) \to \ket{f(t)}}{dt}  = \mathcal N_0 e^{-\Gamma\, t} |A_f|^2 \times & \left\{ \frac{1+|\lambda_f|^2}{2} \cosh \frac{\Delta \Gamma\, t}{2} + \frac{1-|\lambda_f|^2}{2} \cos \Delta m\, t  \right. \nonumber\\
& \left.+ {\rm Re} \lambda_f \sinh \frac{\Delta \Gamma\, t}{2} - {\rm Im} \lambda_f \sin \Delta m\, t \right\}\,,\nonumber \\
\frac{d\Gamma(\ket{\bar B^0(0)}) \to \ket{f(t)}}{dt}  = \mathcal N_0 e^{-\Gamma\, t} |\bar A_f|^2 \times & \left\{ \frac{1+|\bar \lambda_f|^2}{2} \cosh \frac{\Delta \Gamma\, t}{2} + \frac{1-|\bar \lambda_f|^2}{2} \cos \Delta m\, t  \right. \nonumber\\
& \left.+ {\rm Re} \bar \lambda_f \sinh \frac{\Delta \Gamma\, t}{2} - {\rm Im} \bar \lambda_f \sin \Delta m\, t \right\}\,,
\end{align}
where $\mathcal N_0$ is the overall flux normalization,
\begin{align}
\lambda_f &\equiv \frac{q}{p} \frac{\bar A_f}{A_f} \,, & \bar \lambda_f &\equiv \frac{p}{q} \frac{ A_f}{\bar A_f} = \frac{1}{\lambda_f}\,,  
\end{align}
and analogously for decays to $\bar f$\,. The various terms in the above tiem-evolution can be understood as follwing
\begin{itemize}
\item Terms proportional to $|A_f|^2$, $|\bar A_f|^2$ describe a decay without net oscillation.
\item Terms proportional to $|\lambda_f|^2$, $|\bar \lambda_f|^2$ describe a decays following net oscillations.
\item Terms proportional to $\sin \Delta m t$, $\sinh \Delta \Gamma t /2$ describe interference between the above two cases.
\item CP violation in interference is possible only if ${\rm Im} (\lambda_f) \neq 0$\,.
\end{itemize}
Such effects can be observed in neutral $B$ meson decays to CP eigenstates via a time-dependent CP asymmetry
\begin{equation}
A_{f_{CP}} (t) \equiv \frac{\frac{d\Gamma}{dt}\left[\bar B^0(0) \to f_{CP}(t)\right]- \frac{d\Gamma}{dt}\left[ B^0(0) \to f_{CP}(t)\right]}{\frac{d\Gamma}{dt}\left[\bar B^0(0) \to f_{CP}(t)\right] + \frac{d\Gamma}{dt}\left[ B^0(0) \to f_{CP}(t)\right]}\,.
\end{equation}
In the $B$ (and also $B_s$) system experimentally $\Delta \Gamma \ll \Delta m$ and so $|q/p| \simeq 1$\,. In this limit, the full expression for $A_f$ greatly simplifies and can be written as
\begin{equation}
A_f(t) = S_f \sin(\Delta m \,t) - C_f \cos (\Delta m\, t)\,,
\end{equation}
where 
\begin{align}
S_f & \equiv \frac{2\, {\rm Im} (\lambda_f)}{1+|\lambda_f|^2}\,, & C_f & \equiv \frac{1-|\lambda_f|^2}{1+|\lambda_f|^2}\,.
\end{align}

\subsection{Phases in decay amplitudes}
Consider the decay $B \to f$ described by the amplitude $A_f$ and its CP conjugate process $\bar B \to \bar f $ associated with the amplitude $\bar A_{\bar f}$.  
Any complex parameter in the theory Lagrangian entering the two amplitudes will appear complex conjugated after CP and will thus appear with opposite signs in $A_f$ and $\bar A_{\bar f}$\,. The associated CP odd phases are conventionally called {\it weak phases}.  In the SM they are induced via $W$ exchanges. Note that single amplitude phases are convention dependent and thus not physical. Only differences between phases of different amplitudes are physical.

On the other hand, on-shell intermediate states in scattering or decay  ampitudes can produce phase changes even if the relevant Lagrangian is real. These are thus independent of CP. They will appear with same signs in both $A_f$ and $\bar A_{\bar f}$. These CP even phases are often reffered to as {\it strong phases}. In the SM they are due to strong interaction induced re-scattering. Again, only relative phases between amplitudes are physical. 

In general, one can thus write both decay amplitudes as
\begin{align}
A_f & = |a_1| e^{i(\delta_1 +\phi_1)} + |a_2| e^{i(\delta_2 + \phi_2)} + \ldots\,, \nonumber\\
\bar A_{\bar f} & = |a_1| e^{i(\delta_1 -\phi_1)} + |a_2| e^{i(\delta_2 - \phi_2)} + \ldots\,,
\end{align}
where $a_{1,2,\ldots}$ are contributions to the amplitude with different phases, $\delta_{1,2\ldots}$ are the strong phases and $\phi_{1,2\ldots}$ are the weak phases. 

\subsection{CP violation in $B\to \psi K_S$}
To a good approximation, the $B\to \psi K_S$ decays are described by a just single weak decay amplitude to a CP eigenstate (with CP eigenvalue $\eta_f$)
\begin{align}
A_f & = |a_f| e^{i(\delta_f + \phi_f)}\,, \nonumber\\
\bar A_f & = |a_f| e^{i(\delta_f - \phi_f)} \eta_f\,, 
\end{align}
so that $\lambda_f = \eta_f (q/p) \exp(-2i\phi_f)$\,. In the neutral $B$ system $|\Gamma_{12}| \ll |M_{12}|$, since it is due to $\mathcal O(G_F^2)$ long distance effects, which are suppressed by small CKM elements (a fact also verified experimentally since $\Delta \Gamma \ll \Delta m$). Then one can write
\begin{equation}
\left(\frac{q}{p}\right)^2 = \frac{M_{12}^* - \frac{i}{2} \Gamma_{12}^*}{M_{12}-\frac{i}{2} \Gamma_{12}} \simeq e^{2i\xi_B}\,,
\end{equation}
and thus $\lambda_f \simeq \eta_f \exp[i(\xi_B-2\phi_f)]$, leading to a simple expression for the time-dependent CP asymmetry $A_f(t)$, in particular
\begin{equation}
S_{f_{CP}} \simeq \eta_f \sin(\xi_B - 2 \phi_f)\,.
\end{equation}

In the SM, $\xi_B$ and $\phi_f$ are exactly computable in terms of the CKM elements. In particular
\begin{equation}
\xi_B = - {\rm Arg}(M_{12}) \simeq - {\rm Arg}[(V_{tb}^* V_{td})^2] = - {\rm Arg} \left( \frac{V_{tb}^*V_{td}}{V_{tb}V_{td}^*}  \right)\,,
\label{eq:xiB}
\end{equation}
while
\begin{equation}
- e^{-2i\phi_f} = \frac{\bar A^{(B)}_{\psi K_S}}{ A^{(B)}_{\psi K_S}} = - \frac{V_{cb} V_{cs}^*\, a_T + \ldots}{V_{cb}^* V_{cs}\, a_T + \ldots} e^{i \xi_K} \simeq - \frac{V_{cb} V_{cs}^*}{V_{cb}^* V_{cs}} \frac{V^*_{cd} V_{cs}}{V_{cd} V^*_{cs}}\,.
\label{eq:phiB}
\end{equation}
In the above equation, the dots denote additional amplitudes suppressed by small coefficients and CKM elements. Also, in the second step we have taken into account the phase projection of the neutral kaon flavour eigenstates onto the mass eigenstate $K_S$ due to $K - \bar K$ oscillations (analogous to Eq.~\eqref{eq:xiB}). Combining Eqs.~\eqref{eq:xiB} and~\eqref{eq:phiB} we thus obtain
\begin{equation}
\lambda_{\psi K_S}^{(B)} \simeq  \frac{V_{tb}^*V_{td}}{V_{tb}V_{td}^*}  \frac{V_{cb} V_{cd}^*}{V_{cb}^* V_{cd}} = - e^{-2i\beta}\,.
\end{equation}
The observable $S_{\psi K_S}^{(B)} \simeq \sin 2\beta$ (note that $C^{(B)}_{\psi K_S} \simeq 0$) demonstrates {\it CP violation in interference between the mixing and decay} amplitudes. Experimentally, it has been measured to an accuracy of $\sim 1\%$ at the $B$ factories~\cite{pdg}.

\subsection{CP violation in $B_s$ mixing}

Establishing CP violation in the $B_s$ system is considerably more challenging. The golden channel is the decay $B_s \to \psi \phi$\,. Since it is an admixture of different CP eigenestates (represented by the different polarizations of the two vector mesons in the final state), an angular analysis is required for the extraction of the CP violating phase. In addition, $B_s$ oscillations are much faster than those of $B_d$, namely
\begin{equation}
\frac{\Delta m_s}{\Delta m_d} \sim \frac{|M_{12}^s}{|M_{12}^d|} \propto \left| \frac{V_{ts}}{V_{td}} \right|^2 \sim 30\,.
\end{equation}
Finally, $\Delta \Gamma_s$ effects in the time evolution cannot be neglected compared to $\Delta m_s$. In the SM $\lambda_{\psi\phi}^{(B_s)} = - \exp[{i (\xi_{B_s} - 2\phi_{\psi\phi})}]$ evaluates to~\cite{Lenz:2010gu}
\begin{equation}
\left[  S_{\psi\phi}^{(B_s)} \right]_{\rm SM} = 2 {\rm Arg} \frac{V_{tb}^* V_{ts}}{V_{cb}^* V_{cs}} = 0.036(1)\,,
\end{equation}
which is still small compared to the currently attainable experimental precision of $S_{\psi\phi}^{(B_s)} = -0.02(4)$\,~\cite{LHCb-TALK-2014-337}.

\subsection{CP violation in $B$ decays to CP conjugate states}

This form of measurements is interesting if $B^0 \to \bar f$ and $\bar B^0 \to f$ transitions are forbidden. In this case $|A_f| = |\bar A_{\bar f}|$ and $|A_{\bar f}| = |\bar A_f| = 0$ and one can define a CP asymmetry
\begin{equation}
\frac{\frac{d\Gamma}{dt} \left[ \bar B^0(0) \to f(t) \right]  - \frac{d\Gamma}{dt} \left[  B^0(0) \to \bar f(t) \right] }{\frac{d\Gamma}{dt} \left[ \bar B^0(0) \to f(t) \right]  + \frac{d\Gamma}{dt} \left[  B^0(0) \to \bar f(t) \right] } = \frac{\left| \frac{p}{q}\right|^2- \left| \frac{q}{p}\right|^2}{\left| \frac{p}{q}\right|^2+\left| \frac{q}{p}\right|^2} \simeq {\rm Im} \left( \frac{\Gamma_{12}}{M_{12}} \right) + \mathcal O \left( \left| \frac{\Gamma_{12}}{M_{12}}  \right|^2 \right)\,,
\end{equation}
where in the last step we have again used the fact the $|\Gamma_{12}| \ll |M_{12}|$ in the $B$ system. Note that the above asymmetry is accessible with a time-independent measurement. It also represents {\it CP violation in mixing}, simetimes termed {\it indirect CP violation}\,. An illustrative example is the wrong sign semileptonic decay asymmetry
\begin{equation}
a_{SL}^{(d)} = \frac{\Gamma(\bar B^0 \to X \ell^+ \nu) - \Gamma( B^0 \to X \ell^- \bar \nu)}{\Gamma(\bar B^0 \to X \ell^+ \nu)+\Gamma( B^0 \to X \ell^- \bar \nu)}\,,
\end{equation}
with the SM expectation of $a_{SL}^{(d)} = -8(2) \times 10^{-4}$~\cite{Lenz:2010gu}\,.

\subsection{CP violation in charged $B$ decays}
The possibility of CP violation in charged $B$ decays is of special interest in the case of $B^\pm \to D K^\pm$, since $D -\bar D$ oscillations allow for interference of two tree-level dominated decay amplitudes, in particular
\begin{align}
B^- &\to D^0 K^- : ~ b  \to c \bar u s \,, \nonumber\\
B^- &\to \bar D^0 K^- : ~ b  \to \bar c u s \,.
\end{align}
The resulting phenomenology is particularly transparent by focusing on subsequent $D$ decays to CP eigenstates~\cite{Gronau:1990ra}
\begin{align}
D^0 & \to f_{CP} :  ~ c  \to d \bar d u\,, s\bar s u\,, \nonumber\\
\bar D^0 & \to f_{CP} : ~ \bar c  \to d \bar d \bar u\,, s\bar s \bar u\,.
\end{align}
In the SM the ratio of the two decay amplitudes is then
\begin{equation}
\frac{A^{B}_{(D\to f)K}}{A^B_{(\bar D\to f)K}} = \frac{V_{cb}^* V_{us} a^B_{DK}}{V_{ub}^* V_{cs} a^B_{\bar D K}} e^{i(\delta^B_{DK} - \delta^B_{\bar D K})} \eta_f \frac{V_{cd} V_{ud}^*}{V_{cd}^* V_{ud}} \frac{a^D_{f}}{a^{\bar D}_{f}} e^{i(\delta^D_{f}-\delta^{\bar D}_{ f})} \simeq \eta_f r_B e^{i(\delta_B - \gamma)}\,,
\end{equation}
where we have used the definition of the angle $\gamma \equiv {\rm Arg} ( - V_{ud} V_{ub}^* / V_{cd} V_{cb}^* ) \simeq 70^\circ$~\cite{pdg} and have collected the hadronic amplitude ratios into $r_B$ and the associated strong phases in $\delta_B$\,.

The  virtue of these modes is that in principle all unknown parameters can be determined by measuring several available decay rates only, which are CP even quantities. In particular
\begin{align}
A(B^- \to f_+ K^-) &= A_0 \left [ 1 + r_B e^{i(\delta_B - \gamma)} \right]\,, \nonumber\\
A(B^- \to f_- K^-) &= A_0 \left [ 1 - r_B e^{i(\delta_B - \gamma)} \right]\,, \nonumber\\
A(B^+ \to f_+ K^-) &= A_0 \left [ 1 + r_B e^{i(\delta_B + \gamma)} \right]\,, \nonumber\\
A(B^+ \to f_- K^-) &= A_0 \left [ 1 - r_B e^{i(\delta_B + \gamma)} \right]\,.
\end{align}
can be used to extract the three hadronic parameters ($A_0$, $r_B$ and $\delta_B$) as well as $\gamma$\,. Since no $B$ mixing is involved, these measurements are sensitive to {\it CP violation in decay} also termed {\it direct CP violation}. The determination of $\gamma$ in this way is theoretically extremely clean, in particular, since CP violation in $D - \bar D$ mixing is negligible. Experimentally, it is advantageous to have both a large $r_B$ and large $\delta_B$. Therefore, it is welcome that such approach can be adapted also for D-decay products, which are non CP eigenstates~\cite{Atwood:1996ci}.

\section{Flavour and New Physics}

Let us first consider how much NP can still contribute to flavour observables, given the current experimental and theoretical precision. For example, given the good agreement of SM tree-level mediated processes with experiment, one can perform basic tests of CKM unitarity. Taking only the moduli of the first row CKM elements:
\begin{itemize}
\item $|V_{ud}|$ which can be extracted from $0^+ \to 0^+ e \nu$ super-allowed nuclear $\beta$ decays, yielding $|V_{ud}| = 0.97425(22)$~\cite{pdg}\,;
\item $|V_{us}|$ which is determined from the semileptonic kaon decays $K^+ \to \pi^+ \ell \nu$, yielding $|V_{us}| = 0.2237(13)$~\cite{pdg}\,;
\item finally, $|V_{ub}|$ which is measured using charmless semileptonic $B$ decays $B\to X_u \ell \nu$, yelding $|V_{ub}| = 4.2(5) \times 10^{-3}$~\cite{pdg}\,;
\end{itemize}
one can form the following CKM unitarity constraint~\cite{Boyle:2013gsa}
\begin{equation}
|V_{ud}|^2 + |V_{us}|^2 + |V_{ub}|^2 -1 = -0.0008(7)\,.
\end{equation}
Using the measurements of the Fermi constant from the muon life-time, one can further reinterpret these constraints as tests of the charged current universality between leptonic and semileptonic weak processes at the per-mille level. In light of this, it is reasonable to consider NP contributions to observables which are (loop, CKM) suppressed in the SM. Then one can use the CKM determination from tree-level observables, in particular $|V_{ud}|$, $|V_{us}|$, $|V_{cb}|$ and $|V_{ub}|$ as well as $\gamma$ from $B\to D K$ decays (and/or $\alpha$ from tree-level dominated $B\to \pi \pi$ decays)\,. This finally allows to predict SM contributions also to loop suppressed observables, greatly enhancing their sensitivity to NP.

\subsection{New physics in $B - \bar B$ mixing}

In the following we will assume a presence of heavy NP -- such that it would only contribute to dispersive $B - \bar B$ amplitudes. In that case, the most general modification of $M_{12}$ can be parametrised as
\beq
M_{12} = M_{12}^{\rm SM} r_d^2 e^{2i\theta_d}
\eeq
where the NP parameters $r_d$ and $\theta_d$ signify a change of the magnitude and phase with respect to the SM prediction, respectively. Such effects of NP can then be easily translated to all relevant $B$ mixing observables as
\begin{align}
\Delta m_B &= r_d^2 \left( \Delta m_B  \right)^{\rm SM} \,,\nonumber\\
S_{\psi K_S}^{(B)} &\simeq \sin (2\beta + 2\theta_d)\,, \nonumber\\
a_{SL}^{(d)} & = - {\rm Re}\left(  \frac{\Gamma_{12}}{M_{12}} \right)^{\rm SM} \frac{\sin2\theta_d}{r_d^2} + {\rm Im}\left( \frac{\Gamma_{12}}{M_{12}} \right)^{\rm SM} \frac{\cos 2 \theta_d}{r_d^2}\,.
\end{align} 
One can compare these expectations to the current experimental measurements of~\cite{pdg} 
\begin{align}
\Delta m_B & = 51.0(4) \times 10^{10}/s \,, & S_{\psi K_S}^{(B)} & = 0.671(24)\,, & a_{SL}^{(d)} & = -0.2(7) \times 10^{-3}\,,
\end{align}
where the SM expectation with tree-level CKM inputs for $[S_{\Psi K_S}^{(B)}]^{\rm SM}_{\rm tree} = 0.76(4)$~\cite{ckmfitter}\,. We can immediately draw the following conclusions
\begin{itemize}
\item NP in $M_{12}$ with a large phase relative to $\beta$ is constrained to $20\% - 30\%$ of the SM contribution. Thus, CKM clearly dominates CP violation in $B - \bar B$ mixing. (A similar conclusion can be made for the case of $K^0 - \bar K^0$ system: the measured value of $\epsilon_K = 1.596(13)\times 10^{-3}$ constrains CP violating NP in kaon mixing amplitudes to be subdominant. The power of this constraint is however presently limited by theory uncertainties.)
\item NP in $M_{12}$ with a phase that is aligned to $\beta$ is constrained to be at most comparable to the SM contribution. (Again a similar conclusion can be made for the case of $K^0 - \bar K^0$ system: the measured value of $\Delta m_K = 52.93(9)\times 10^{8}/s$ constrains CP conserving NP in kaon mixing amplitudes to be comparable to SM estimates, which however contrary to the $B$ case carry sizable theory uncertainties.)
\end{itemize}
As a comparison, in the case of $B_s$ mixing, NP can be at most comparable to the SM contribution regardless of the phase since $S_{\psi\phi}^{\rm SM} \lesssim \delta S_{\psi\phi}^{\rm exp}$\,.

\subsection{The new physics flavour puzzle}

The SM is not a complete theory of Nature.
\begin{enumerate}
\item It does not include a (quantum) description of gravity. Thus its validity is limited below the Planck scale $m_{\rm Planck} \simeq 10^{19}$~GeV.
\item It does not include neutrino masses. This further limits its validity down to below the maximal scale at which new degrees of freedom can accommodate at least two massive neutrinos $m_{\rm see-saw}\lesssim 10^{15}$~GeV.
\item The fine-tuning of the EW symmetry breaking scale compared to the large scales in the above points 1. and 2. suggests NP already at scales of the order $4\pi v \sim 1$~TeV\,.\footnote{Incidentally, the TeV mass scale can also be associated with the explanation of the cosmological dark matter, if the later is in the form of a thermal particle relic.}
\end{enumerate}
Given the SM is merely an effective field theory valid below a cut-off energy scale $\Lambda$, one needs to consider additional terms in the theory Lagrangian consisting of SM field operators with canonical dimensions $d>4$:
\beq
\mathcal L = \mathcal L_{\rm SM} + \sum_{d>4} \sum_n \frac{c_n^{(d)}}{\Lambda^{d-4}} \mathcal O_n^{(d)}\,.
\eeq
In a natural theory one expects $c_n^{(d)} \sim \mathcal O(1)$ unless the relevant operators are forbidden or suppressed by symmetries. For $\Lambda \sim $~TeV and without imposing additional symmetries beyond the gauged SM ones, the above condition is severely violated for several $\mathcal O_n^{(6)}$\,, which contribute to flavour changing processes. This constitutes the so-called {\it NP flavour puzzle}\,, which can be articulated through the following question: If there is NP at the TeV scale, why haven't we seen its effects in flavour observables? Naively, one could argue, that the same it true for NP violating baryon and lepton numbers. However, $B$ and $L$ are (classically) exact accidental symmetries of the SM, while in the SM the flavour symmetry is already broken explicitly. 

\subsection{Bounds on new physics from $\Delta F=2$ processes}
The NP flavour puzzle can be demonstrated perhaps most dramatically in the case $\Delta F=2$ FCNCs. In the SM the dispersive contributions to $\Delta F=2$ processes of down-quarks are typically dominated by box diagrams with the top quarks appearing in the loop. These contributions can be schematically written as
\beq
M_{12}^{\rm SM} = \frac{G_F^2 m_t^2}{16\pi^2} \left( V_{ti}^* V_{tj} \right)^2 \bra{\bar M} (\bar d_L^i \gamma_\mu d_L^j)^2 \ket{M} F\left( \frac{m_t^2}{m_W^2} \right) + \ldots\,,
\eeq
 where $M=K^0,~B^0,~B_s$, $d^{i,j}$ denote meson valence quarks, $F(x) \sim \mathcal O(1)$ is the relevant loop function normalized to $F(\infty) = 1$\,, while the dots denote corrections due to charm quark contributions, which are numerically relevant only in the case of $K - \bar K$ mixing. Note that the prefactor can be rewritten completely in terms of the fundamental flavour parameters (Yukawas) in the unbroken theory
\beq
\frac{G_F^2 m_t^2}{16\pi^2} \left( V_{ti}^* V_{tj} \right)^2 = \frac{(Y_u Y_u^*)_{ij}}{128\pi^2 m_t^2}\,,
\eeq 
which can be interpreted as due to Goldstone Higgs exchanges in the gaugeless $(g\to 0)$ limit of the SM.

The relevant hadronic matrix elements between the external $M$ and $\bar M$ mesons can be written as
\beq
\bra{\bar M} (\bar d^i_L \gamma_\mu d_L^j) (\bar d^i_L \gamma^\mu d_L^j) \ket{M} = \frac{2}{3} f_M^2 m_M^2 \hat B_M\,,
\eeq
where the relevant meson decay constant $f_M$ is defined via $\bra{0} d^i \gamma_\mu\gamma_5 d^j \ket{M (p)} \equiv i p_\mu f_M$\,, while $\hat B_M \sim \mathcal O(1)$ is called the bag parameter.  These two hadronic quantities can be computed numerically using lattice QCD methods.The tremendous progress in these calculations over the past 30 years is reflected in the precise values of~\cite{flag}
\begin{align}
f_B & = 0.186(4)~{\rm GeV}\,, & \hat B_B & = 1.27(10)\,, \nonumber\\
f_{B_s} & = 0.224(5)~{\rm GeV}\,, & \hat B_{B_s} & = 1.33(6)\,, \nonumber\\
f_K & = 0.1563(9)~{\rm GeV}\,, & \hat B_K & = 0.7661(99)\,. 
\end{align}
 With these inputs we can use the experimental measurements of neutral meson mixing observables to constrain possible NP contributions of the form
 \begin{align}
 \mathcal L_{\rm NP}^{\Delta F=2} & = \frac{c_{sd}}{\Lambda^2} \left( \bar d_L \gamma^\mu s_L \right)^2 + \frac{c_{bd}}{\Lambda^2} \left( \bar b_L \gamma^\mu s_L \right)^2 + \frac{c_{bs}}{\Lambda^2} \left( \bar s_L \gamma^\mu b_L \right)^2 \nonumber\\
 &+ \frac{c_{cu}}{\Lambda^2} \left( \bar u_L \gamma^\mu c_L \right)^2 + \frac{c_{tu}}{\Lambda^2} \left( \bar u_L \gamma^\mu t_L \right)^2 + \frac{c_{tc}}{\Lambda^2} \left( \bar c_L \gamma^\mu t_L \right)^2\,.
 \end{align}
The effects of such NP on neutral meson oscillations can namely be completely encoded into
\beq
\frac{M_{12}^M}{m_M} \sim c_{ij} \left(\frac{f_M}{\Lambda}\right)^2\,,
\eeq
which leads to the following set of current experimental constraints~\cite{utfit,df=2}
\begin{align}
\frac{\Delta m_K}{m_K} &\sim 7 \times 10^{-15} & \Rightarrow && \frac{\Lambda}{\sqrt{|c_{sd}|}} & \gtrsim 10^{3}~{\rm TeV}  & {\rm or} & & |c_{sd}| & \lesssim 10^{-6} \left( \frac{\Lambda}{\rm TeV} \right)^2\,, \nonumber\\
\frac{\Delta m_D}{m_D} &\sim 9 \times 10^{-15} & \Rightarrow && \frac{\Lambda}{\sqrt{|c_{cu}|}} & \gtrsim 10^{3}~{\rm TeV}  & {\rm or} & & |c_{cu}| & \lesssim 10^{-6} \left( \frac{\Lambda}{\rm TeV} \right)^2\,, \nonumber\\
\frac{\Delta m_B}{m_B} &\sim 6 \times 10^{-14} & \Rightarrow && \frac{\Lambda}{\sqrt{|c_{bd}|}} & \gtrsim 4\times 10^{2}~{\rm TeV}  & {\rm or} & & |c_{bd}| & \lesssim 5\times 10^{-6} \left( \frac{\Lambda}{\rm TeV} \right)^2\,, \nonumber\\
\frac{\Delta m_{B_s}}{m_{B_s}} &\sim 2 \times 10^{-12} & \Rightarrow && \frac{\Lambda}{\sqrt{|c_{bs}|}} & \gtrsim 70~{\rm TeV}  & {\rm or} & & |c_{bs}| & \lesssim 2\times 10^{-4} \left( \frac{\Lambda}{\rm TeV} \right)^2\,.
\end{align} 
Furthermore, in case of maximal CP violating phases in $c_{ij}$, one obtains even stronger constraints
\begin{align}
\epsilon_K &\sim 0.0023 & \Rightarrow && \frac{\Lambda}{\sqrt{|{\rm Im}(c_{sd})|}} & \gtrsim 2 \times 10^{4}~{\rm TeV}  & {\rm or} & & |{\rm Im} (c_{sd})| & \lesssim 6\times 10^{-10} \left( \frac{\Lambda}{\rm TeV} \right)^2\,, \nonumber\\
\frac{A_\Gamma}{y_{CP}} &\lesssim 0.2 & \Rightarrow && \frac{\Lambda}{\sqrt{|{\rm Im}(c_{cu})|}} & \gtrsim 3 \times 10^{3}~{\rm TeV}  & {\rm or} & & |{\rm Im} (c_{cu})| & \lesssim 10^{-7} \left( \frac{\Lambda}{\rm TeV} \right)^2\,, \nonumber\\
S_{\psi K_S} &\sim 0.67 & \Rightarrow && \frac{\Lambda}{\sqrt{|{\rm Im}(c_{bd})|}} & \gtrsim 8 \times 10^{2}~{\rm TeV}  & {\rm or} & & |{\rm Im} (c_{bd})| & \lesssim 10^{-6} \left( \frac{\Lambda}{\rm TeV} \right)^2\,, \nonumber\\
S_{\psi\phi} &\sim 0.1 & \Rightarrow && \frac{\Lambda}{\sqrt{|{\rm Im}(c_{bs})|}} & \gtrsim 70~{\rm TeV}  & {\rm or} & & |{\rm Im} (c_{bs})| & \lesssim 2\times 10^{-4} \left( \frac{\Lambda}{\rm TeV} \right)^2\,.
\end{align} 
 
The two main messages one can draw from such an analysis are that (1) NP with a generic flavour structure is irrelevant for EW hierarchy, since flavour measurements in this case require $\Lambda \gg $~TeV; and (2) in case of TeV NP, its flavour structure  needs to be far from generic.

\section{Conclusions}
The absence of significant deviations from the SM in quark flavour physics is a key constraint on
any extension of the SM. At the same time there are still various open questions regarding the flavour structure of the standard model itself that can be possibly addressed only at low energies, using flavour phyiscs measurements. The set of flavour observables to be measured with higher precision in the search for indirect hints of NP is limited, but not necessarily small. For example, we still  have only limited knowledge about CP violation in the
$B_s$ and $D$ systems. In addition,  despite significant recent progress, new-physics effects could still be hidden in certain rare kaon, $D$ and $B$ decays~\cite{Buchalla:2008jp}. The experimental progress on these, as expected from the LHCb~\cite{Bediaga:2012py} in LHC run II,  Belle II~\cite{Aushev:2010bq} and other upcoming flavour experiments will thus be invaluable.
 
\subsection*{Acknowledgements}
I wish to thank the organizers of the CERN's 2014 European School of High-Energy Physics for the invitation to lecture in this remarkable setting. I am also grateful to the students and discussion leaders for the stimulating
questions and discussions.

\end{document}